\documentclass[%
 superscriptaddress,
 preprint,
 showpacs,preprintnumbers,
 amsmath,amssymb,
 aps,
 prb,
]{revtex4-2}

\usepackage{graphicx}% Include figure files
\usepackage{dcolumn}% Align table columns on decimal point
\usepackage{bm}% bold math
\usepackage{upgreek}
\usepackage{amsmath}

\begin{document}

%\preprint{APS/123-QED}

\title{Efficient photon-pair generation empowered by dual quasi-bound states in the continuum}% Force line breaks with \\

\author{Tingting Liu}
\affiliation{Institute for Advanced Study, Nanchang University, Nanchang 330031, China}

\author{Meibao Qin}
\affiliation{School of Education, Nanchang Institute of Science and Technology, Nanchang 330108, China}

\author{Siqi Feng}
\affiliation{Institute for Advanced Study, Nanchang University, Nanchang 330031, China}

\author{Xu Tu}
\affiliation{Institute for Advanced Study, Nanchang University, Nanchang 330031, China}

\author{Tianjing Guo}
\affiliation{Institute of Space Science and Technology, Nanchang University, Nanchang 330031, China}

\author{Feng Wu}
\affiliation{School of Optoelectronic Engineering, Guangdong Polytechnic Normal University, Guangzhou, 510665, China}

\author{Shuyuan Xiao}
\email{syxiao@ncu.edu.cn}
\affiliation{Institute for Advanced Study, Nanchang University, Nanchang 330031, China}

\begin{abstract}
	
Here we demonstrate the efficient photon-pair generation via spontaneous parametric down conversion from a semiconductor metasurface supporting dual quasi-bound states in the continuum (quasi-BICs). In a simple metasurface design composed of AlGaAs ellipse nano-cyclinders, the two high-$Q$ quasi-BIC resonances that coincide with the generated signal and idler frequencies significantly boost the local electric field. This leads to a substantial enhancement in the reverse classical nonlinear process of sum frequency generation and subsequently the remarkable high generation rate of photon pairs under the quantum-classical correspondence principle. Within a narrowband wavelength regime around the quasi-BIC resonances, the rate of pair production is enhanced up to $\sim10^{4}$ Hz, two orders of magnitude larger than that in the Mie resonant AlGaAs nanoantennas. Moreover, the photon pair emission is mainly concentrated in the normal direction with respect to the metasurface, and shows tunable rate with the $Q$ factor by engineering the rotation angle of nano-cylinders. The presented work enables nanoscale sources of high-quality entangled photons which will find applications in advanced quantum imaging and communications.

\end{abstract}

%\pacs{42.70.-a, 42.79.-e, 78.67.Pt}% PACS, the Physics and Astronomy
\maketitle

%\tableofcontents

\section{\label{sec1}Introduction}

The ability to generate and manipulate complex optical photon states involving entanglement between multiple optical modes is not only the fundamental key to advancing photonic quantum technologies, but will revolutionize a wide range of applications such as quantum communication, computation, imaging, and microscopy\cite{MaganaLoaiza2019, Wang2020}. The spontaneous parametric down-conversion (SPDC) is arguably the most versatile source of heralded single photons and entangled photon pairs, owing to high indistinguishability, room-temperature operation, simple signal filtering, and coherent emission as well as entanglement in several degrees of freedom. Despite these advantages, the stringent phase matching and the significant volume footprint have been major challenges facing current research. One main path to lifting these limitations is the miniaturization of SPDC-based photon sources to the sizes smaller than the coherence length. The earlier approaches were demonstrated in ultrathin films of nonlinear materials\cite{Okoth2019, SantiagoCruz2021, Sultanov2022}. However, the achieved photon-pair generation rates were modest due to the extremely weak parametric amplification of the vacuum field. Recently, strong resonant modes in the all-dielectric nanostructures have been prompted to boost the nonlinear light-matter interactions, bringing advances to the generation of photon pairs\cite{Solntsev2021, Grinblat2021, Liu2022, Sharapova2023}. Several pioneering works reported that in such compact nanostructures, for example nanoantennas\cite{Marino2019, Nikolaeva2021, Saerens2023, Weissflog2024} and metasurfaces\cite{SantiagoCruz2021a}, the enhanced SPDC process was realized for quantum state generation and effective manipulation of the quantum features of emitted photons, via taking advantage of the significant field confinement from Mie resonances. In these studies, the $Q$ factor of the resonances related to the capability of local electromagnetic field confinement, becomes a principal indicator in enhancement of nanoscale nonlinear interactions. This leads to significant research interests in high-$Q$ factor resonances for nanoscale photon-pair sources.
 
Most recently, ultrahigh-$Q$ factor resonances based on bound states in the continuum (BICs) are suggested to facilitate the light-matter interaction in dielectric nanostructures\cite{Hsu2016, Sadreev2021, Huang2023}. Optical BICs are the peculiar discrete-energy modes embedded to the radiation continuum but remain perfectly confined without any radiation. True BICs exist in ideal lossless infinite structures, exhibiting an infinitely high-$Q$ factor and zero resonant width, while they can be transformed into quasi-BICs by breaking symmetries such as adding or removing part of nano-resonators\cite{Koshelev2018, Cong2019, Li2019, Wang2020a, Qin2022}, or by adjusting the geometrical parameters including the period, gap, and rotation angle\cite{He2018, Wu2019, Tian2020, Wang2023, Sun2023}. In practical, the quasi-BIC resonances with remarkable local field confinement have been demonstrated to enhance the efficiency of harmonic generation\cite{Carletti2018, Xu2019, Koshelev2019, Carletti2019, Liu2019, Koshelev2020, Ning2021, Zograf2022, Xiao2022, Huang2022, Zalogina2023, Liu2023a},  and wave mixing processes\cite{CamachoMorales2022, Xu2022, Sun2022, Liu2023, Feng2023, Cai2023}, by several orders of magnitude. Due to the similarities between classical parametric frequency conversion and SPDC, comparable enhancements for SPDC can be expected. It has been suggested a single quasi-BIC resonance enhances the quantum vacuum field at the signal wavelength, which in turn, increases the rate of SPDC\cite{Son2023}. In further exploration, metasurfaces with quasi-BIC resonances at both the signal and idler frequencies can be employed to enhance the photon-pair generation rate and spectral brightness. This approach has been theoretically and experimentally studied in recent reports of symmetry-protected BIC metasurfaces, such as metasurfaces composed of a square array of AlGaAs cylinders with two holes\cite{Parry2021}, or GaAs cubes with a small notch\cite{Santiago-Cruz2022}. The photon emission was greatly enhanced by quasi-BICs in such metasurfaces with reduced-symmetry nanoresonators, however, symmetry breaking by adding or removing part of nano-resonators inevitably burdens the fabrication process, and the fabrication imperfections may detrimentally affect quantum fidelity.

In this work, we resolve these problems by designing a dual quasi-BIC metasurface consisting of elliptical nano-cylinders to efficiently generate correlated photon pairs via SPDC. In the designed metasurface, the two quasi-BIC resonances at signal and idler in the telecommunication band around 1550 nm wavelength generate extremely strong electric field enhancement around the edges or within the nonlinear dielectric cylinders. Benefiting from the enhanced electric energy, the reverse classical nonlinear process, sum-frequency generation (SFG) is substantially increased with nonlinear efficiency up to $2.64\times10^{-3}$, and the pair generation rate calculated from the quantum-classical correspondence relation is substantially increased by two orders of magnitude larger compared to the Mie resonant AlGaAs nanoantennas. Due to the nonlocal ultrahigh-Q feature of quasi-BIC resonances, the photon pairs are emitted only in a narrow wavelength range. By engineering the rotation angles of elliptical nano-cylinders in the proposed metasurface, we demonstrate the tunable SPDC efficiency arising from the changing $Q$ factors. Its efficient SPDC response, simple and compact design are beneficial for various emerging free-space quantum optical devices. These results open the way to build nanoscale sources for high-quality entangled photon pairs and can find a plethora of interesting applications in advanced quantum imaging and communications.
 
\section{\label{sec2}Dual quasi-BICs in all-dielectric metasurfaces}

We consider the SPDC photon-pair source of an AlGaAs metasurface supporting dual quasi-BIC resonances at signal and idler frequencies. In the SPDC process, a higher-energy pump photon at $\omega_{\text{p}}$ spontaneously decays due to the strong second-order nonlinear susceptibility $\chi^{(2)}$, and downconverts into two correlated and entangled photons called signal and idler at the angular frequencies $\omega_{\text{s}}$ and $\omega_{\text{i}}$, respectively. This process is schematically shown in Fig. 1(a). Restricted from the typical limitation of phase-matching condition for nonlinear optical process, the energy conservation of $\omega_{\text{p}}=\omega_{\text{s}}+\omega_{\text{i}}$ is always satisfied, while the ultrathin source of metasurfaces leads to the longitudinal momentum conservation relaxed. Since the characterization of SPDC process requires random variables with complex computation, we first use the metasurface in the reverse classical nonlinear SFG process, and then investigate the SPDC process based on the quantum-classical correspondence between them. In SFG process, the two photons with frequencies $\omega_{1}$ and $\omega_{2}$ merge to become a higher-energy photon at $\omega_{\text{SFG}}$, which is in the reverse direction to the SPDC process with a single pump photon splitting into two, as sketched in Fig. 1(a). Accordingly, the efficiency of SPDC generated photon pairs can be predicted by studying classical SFG process under reversed direction interacting waves.

\begin{figure*}[htbp]
\centering
\includegraphics% Here is how to import EPS art
[scale=0.25]{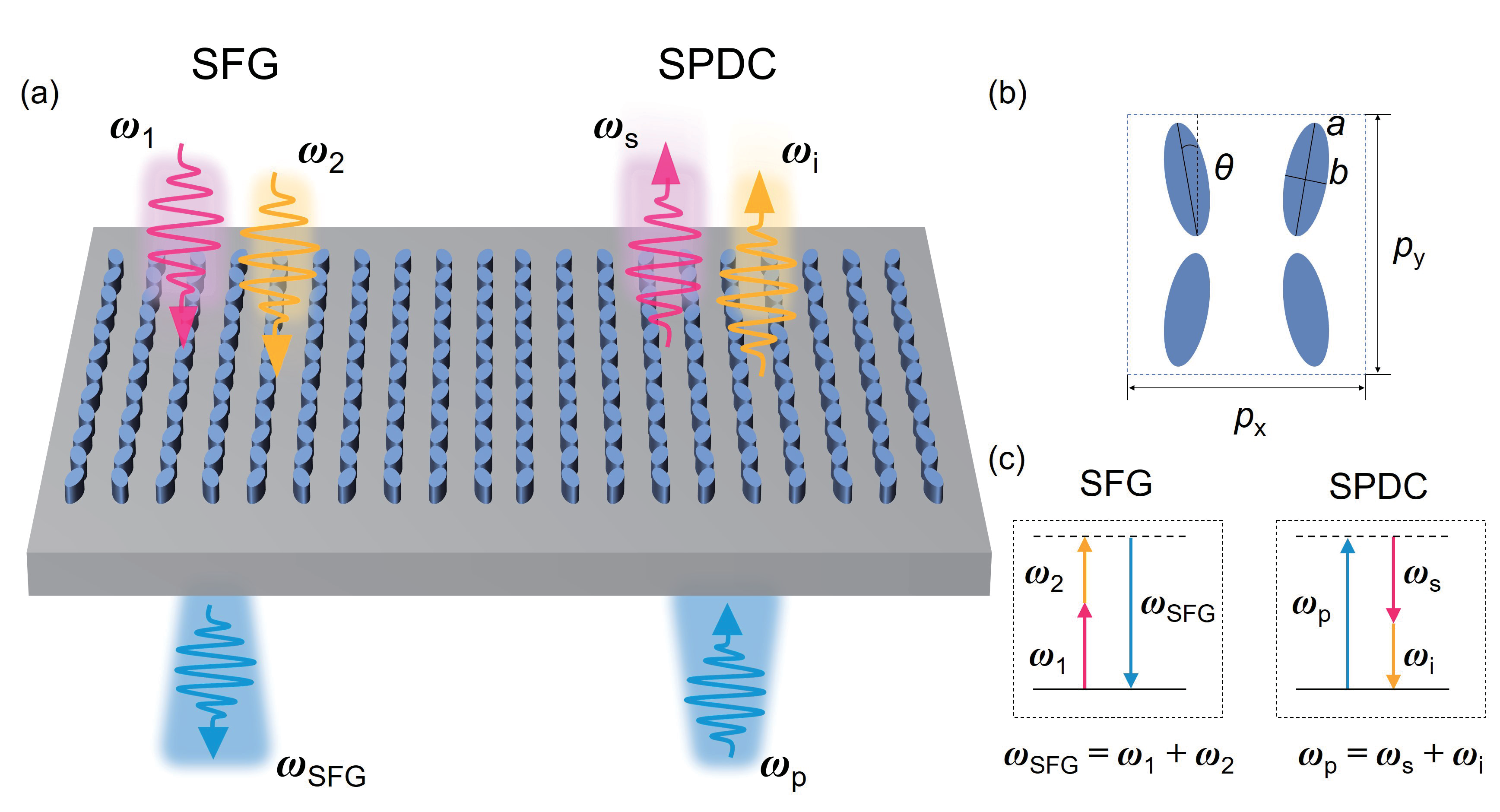}
\caption{\label{fig1} Dual quasi-BIC metasurface for nonlinear processes. (a) The schematic of the designed metasurface for SPDC and its reverse SFG process. (b) The unit cell consists of four elliptical nano-cylinders. (c) The energy diagram of SFG and SPDC process.}
\end{figure*}

We use a specific design of nonlinear all-dielectric metasurface which comprises four AlGaAs elliptical nano-cylinders in each unit cell, as shown in Fig. 1(b). The corresponding lattice constant is denoted by $p_{x}=1240$ nm and $p_{y}=1360$ nm. The parameters of the elliptical nano-cylinders are long axis $a=600$ nm, short axis $b=220$ nm, and height $h=360$ nm, respectively. The nano-resonators are positioned on top of an AlOx substrate. The material of AlGaAs is well known for its strong second-order nonlinear susceptibility $\chi^{(2)}$, and it poses an excellent alternative for nonlinear interaction in metasurface platform operating in the telecommunication wavelength of interest\cite{Carletti2018, Rocco2018, Marino2019, Liu2021, Zalogina2022, Zalogina2023}. The meta-atom design in the form of elliptical nano-cylinders has been experimentally demonstrated in several works\cite{Tittl2018, Yesilkoy2019, Anthur2020, Zhang2022, Dong2022, Son2023}, showing the efficiency in enhanced nonlinear processes and reliability in fabrication and experimental setup. This AlGaAs metasurface is designed so that it supports two high-$Q$ quasi-BIC resonances. The orientation of each cylinder is denoted by rotation angle $\theta$ to constitute the adjusting parameter. When the cylinders are parallel with $\theta=0^{\circ}$, the metasurface supports infinite high-$Q$ eigenmodes at the $\Gamma$ point at the wavelengths of interest, due to the completely decoupling from the far field, as shown in Fig. S1(a). By tilting the elliptical nano-cylinders with different theta, the coupling can be controlled and the $Q$ factors of the resonances can be tuned. Fig. S1(b) schematically depicts the evolution process of the two quasi-BICs with reducing $Q$ factor as the larger rotation angle $\theta$ enables the stronger mode coupling to the far field. 

We identify the two quasi-BIC resonances in the near-infrared spectral region in Fig. 2(a). As $\theta$ increases from $0^{\circ}$ to $15^{\circ}$, two distinct resonance dips can be observed under normal incident illumination. For the metasurface with parallel cylinders, i.e. $\theta=0^{\circ}$, the two resonance width vanishes with an infinite $Q$ factor. The metasurfaces with increasing $\theta$ exhibit the gradually broaden resonance widths, because of the increasing coupling to far field. This tendency agrees well with the evolution of $Q$ factor in Fig. S1(b). The transmission spectrum in the metasurface with $\theta=10^{\circ}$ is exemplified in Fig. 2(b). The transmission dips due to the dual quasi-BIC resonances are observed at 1471 nm (Q-BIC 1) and 1571 nm (Q-BIC 2), respectively. The simulated spectrum can be fitted with the Fano formula $T=|a_{1}+ia_{2}+\frac{b}{\omega-\omega_{0}+i\gamma}|^{2}$, where $a_{1}$, $a_{2}$, and $b$ are real numbers, $\omega_{0}$ is the resonance frequency, and $\gamma$ is the leakage rate\cite{Xu2019, Li2019, Wang2020a, Xiao2022, Liu2023}. Derived from these, the $Q$ factors of the two resonances are calculated by $\omega_{0}/2\gamma$ as 856 and 440 corresponding to Q-BIC 1 and Q-BIC 2, respectively. The two high-$Q$ quasi-BICs provide large field enhancement of the photon-pair emission. As shown in the inset of Fig. 2(b), the electric field enhancements are around the edges of AlGaAs cylinders for Q-BIC 1 or inside the cylinders for Q-BIC 2. To characterize the ability of quasi-BIC resonances to maximize the excitation density within nonlinear materials, we analyze the electric energy stored inside the AlGaAs resonators by integrating the electric energy over the cylinder volume. In Fig. S2, the two maxima can be observed at resonant wavelengths. 

\begin{figure*}[htbp]
\centering
\includegraphics% Here is how to import EPS art
[scale=0.4]{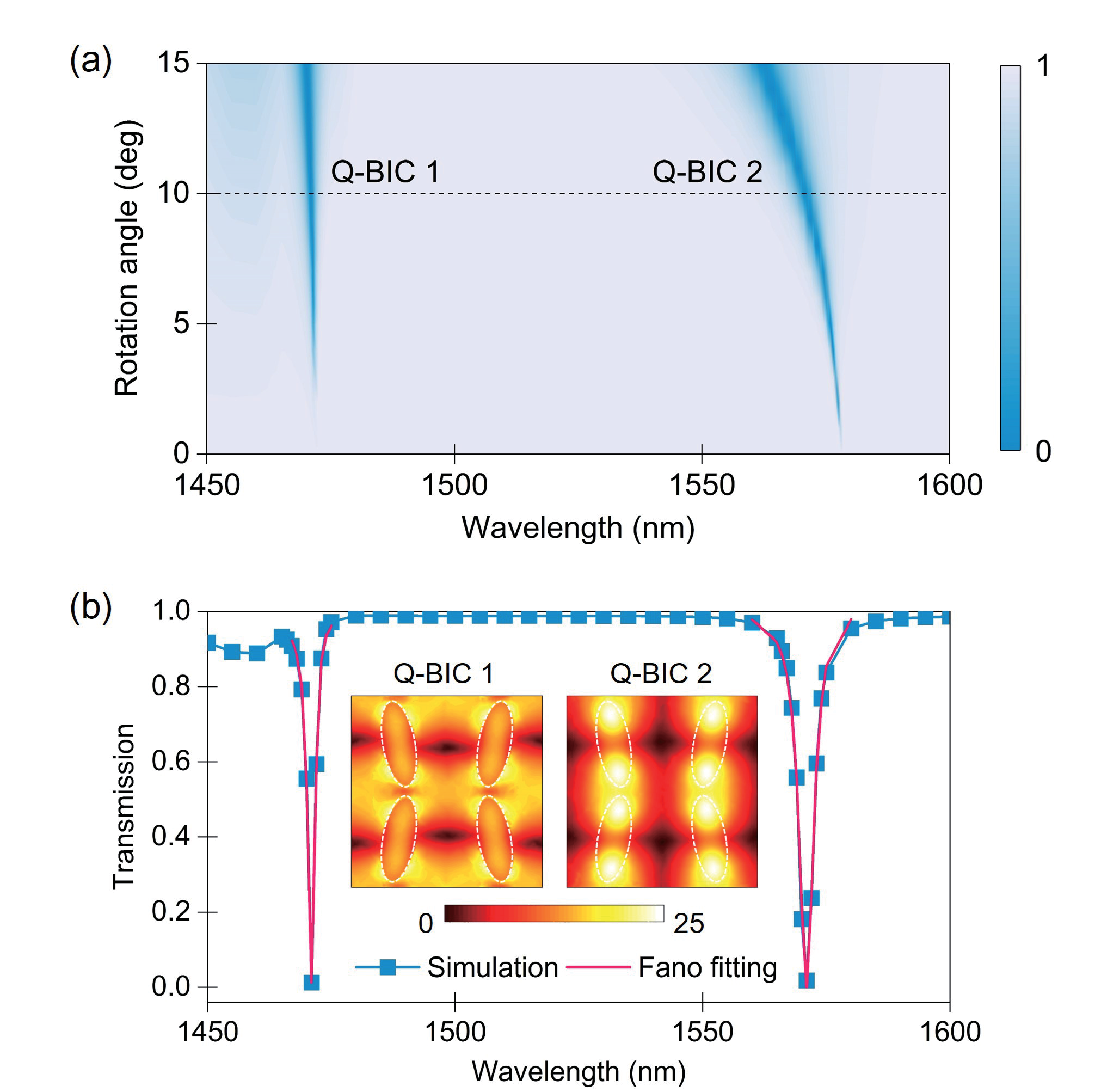}
\caption{\label{fig2} Transmission spectra of the dual quasi-BIC metasurface under normal incidence of $x$-polarized plane waves propagating along the $z$-axis. (a) The transmission spectra as a function of rotation angle of the elliptical nano-cylinders. (b) The transmission spectrum around the two quasi-BIC resonances with rotation angle of $10^{\circ}$. The insets are the electric field distributions at the resonant wavelengths.}
\end{figure*}

\section{\label{sec3}Quantum-classical correspondence between SPDC and SFG}

To gain a quantitative estimation of the quantum photon-pair rate generated through SPDC process, we first consider its classical reverse process SFG. For the SFG process here, the two plane waves and illuminate from air and generate SFG signal ($\omega_{\text{SFG}}=\omega_{1}+\omega_{2}$) propagating in the substrate, all in the opposite directions to the signal ($\omega_{\text{s}}$), idler ($\omega_{\text{i}}$), and pump($\omega_{\text{p}}$) photons in the SPDC process sketched in Fig. 1(a). The quantum-classical correspondence between SPDC and SFG has been performed for arbitrary nonlinear structures with quadratic nonlinearity based on the general Green’s function formalism\cite{Poddubny2016}. It allows a mathematically equivalent approach to predict the SPDC efficiency in a more convenient approach of numerical modeling, instead of the complicated quantum calculations. Based on the quantum-classical correspondence, the quantum photon-pair rate of the SPDC process is derived by the classical reverse SFG process as\cite{Marino2019, Jin2021, Parry2021, Mazzanti2022, Zhang2022a},
\begin{equation}
	\frac{1}{\Phi_{\text{p}}}\frac{dN_{\text{pair}}}{dt}=2\pi\Xi_{\text{SFG}} \frac{\lambda_{\text{p}}^{4}}{\lambda_{\text{s}}^{3}\lambda_{\text{i}}^{3}}\frac{c\Delta\lambda}{\lambda_{\text{s}}^{2}},\label{eq1}
\end{equation}
where $\Phi_{\text{p}}$ is the SPDC pump flux, $c$ is the light speed in vacuum, $\lambda_{\text{p}}$, $\lambda_{\text{s}}$, and $\lambda_{\text{i}}$ denote the pump, signal, and idler wavelengths, respectively. $\Delta\lambda$ is the nonlinear resonance bandwidth at the signal/idler wavelengths. Considering SPDC is produced from the spontaneous parametric amplification of vacuum thermal noise photons, occurrence of all the combinations of $\lambda_{\text{s}}$ and $\lambda_{\text{i}}$ satisfying the energy conservation condition can be possible from the quantum vacuum fluctuations. Thus the bandwidth $\Delta\lambda$ is primarily dependent on the spectral response of the structure and especially the bandwidth of the detector. The normalized efficiency $\Xi_{\text{SFG}}$ is defined as the ratio of the corresponding SFG output power to the product of the incident intensities of the fundamental pumps at the signal and idler frequencies, namely, $\Xi_{\text{SFG}}=P_{\text{SFG}}/I_{1}/I_{2}$. According to the quantum-classical correspondence, the photon-pair rate of SPDC in the designed resonant metasurface is proportional to the normalized efficiency of reverse SFG process.

In this framework, the SFG process is firstly investigated for predicting the photon-pair generation rate of SPDC. The classical SFG process is simulated using the coupled electromagnetic waves frequency domain interfaces in COMSOL Multiphysics based on the finite element method solver. The first and second electromagnetic models are simulated at the fundamental wavelengths $\lambda_{1}$ and $\lambda_{2}$ to retrieve the local field distributions and compute the nonlinear polarization induced inside the nonlinear material, then this polarization term is employed as the only source for the third electromagnetic model at the sum-frequency wavelength to obtain the generated photon power flux propagated to the substrate. In the AlGaAs material, the nonlinear polarization term $P^{\text{NL}}$ is given by 
\begin{equation}
	 P_i^{\text{NL}}(\omega_{\text{SFG}})=\varepsilon_{0}\chi_{ijk}^{(2)}[E_{j}(\omega_{1})E_{k}(\omega_{2})+E_{k}(\omega_{1})E_{j}(\omega_{2})],\label{eq2}
\end{equation}
where $\varepsilon_{0}$ is permittivity of free space, $i$, $j$, and $k$ represent the Cartesian components $x$, $y$, and $z$, respectively. Due to the zincblende lattice, AlGaAs has the only nonvanishing terms of the second-order nonlinear susceptibility tensor $\chi_{ijk}^{(2)}$ with $i\neq j\neq k$. Following the experimental reports\cite{Ohashi1993}, we set all components of the $\chi^{(2)}$ to 100 pm/V. Here the fundamental pump intensities are initially set to 1 GW/cm$^{2}$. The SFG power is computed by integrating the Poynting vector over the output plane on the transmission side. 

Figure 3 shows the simulated conversion efficiency of SFG process in the proposed AlGaAs metasurface supporting dual quasi-BIC resonances. The conversion efficiency is defined as the output power of generated SFG signal normalized by the sum of input powers of the two fundamental waves, $\eta=P_{\text{SFG}}/(P_{1}+P_{2})$. As shown in Fig. 3(a), when the input fundamental waves with wavelengths are close to the quasi-BIC resonances, the SFG process is remarkably enhanced. As the fundamental wavelengths match with the two quasi-BIC resonances at 1471 nm and 1571 nm respectively, the largest efficiency reaches $\eta=2.64\times10^{-3}$. Meanwhile, $\eta$ shows strong dependence on the input wavelength, manifested by a narrow range of boosted SFG efficiency within 2 nm. This feature of narrowband enhancement arises from nonlocal ultrahigh-$Q$ resonance nature of the quasi-BICs that exhibit narrow resonant widths as depicted in Fig. 2(b). To elaborate the resonant enhancement effect of the quasi-BICs in SFG process, the simulated $\eta$ around the two quasi-BIC resonances are provided in Figs. 3(c) and 3(d), respectively. When one of the input beam is fixed at the resonance wavelength 1571 nm of the Q-BIC 2, the conversion efficiency of SFG signal reaches the maximum at the Q-BIC 1 resonance with wavelength 1471 nm. The peak efficiency can also be observed for the similar situation in Fig. 3(d). It can be concluded that the maximum enhancement of the SFG process only appears at the fundamental wavelengths corresponding to the two quasi-BIC resonances, confirming the collaboratively increased SFG efficiency from resonant enhancement of dual quasi-BICs at the two fundamental wavelengths. Meanwhile, the linear dependence of the SFG power on the fundamental power can be derived from Fig. 3(b). It suggests that the nonlinear SFG process can be further boosted by modifying the input power, for example, the efficiency up to $\sim3\times10^{-2}$ can be achieved by increasing input power at the two fundamental wavelengths to 10 GW/cm$^{2}$.

\begin{figure*}[htbp]
	\centering
	\includegraphics% Here is how to import EPS art
	[scale=0.35]{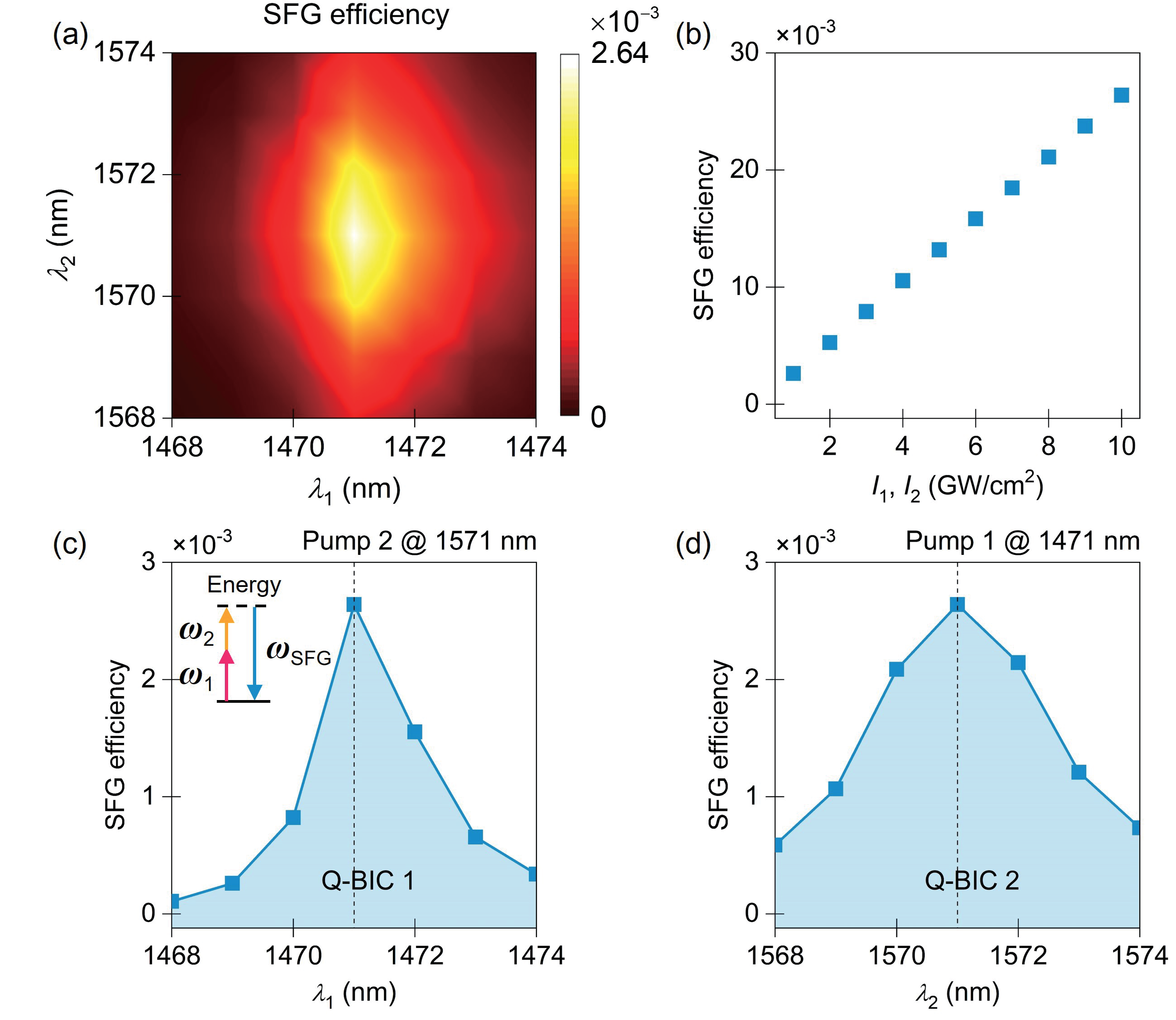}
	\caption{\label{fig3} Enhanced SFG efficiency in the dual quasi-BIC metasurface. The SFG conversion efficiency as a function of (a) the pump wave incident wavelengths around the resonances, (b) the input intensity of the pump waves, (c) the pump wavelength $\lambda_{1}$ for the case of $\lambda_{2}$ fixed at Q-BIC 2, and (d) the pump wavelength $\lambda_{2}$ for the case of $\lambda_{1}$ fixed at Q-BIC 1.}
\end{figure*}

Based on the quantum-classical correspondence, we can predict the efficiency of quantum photon-pair generation of SPDC process. For calculations, we assume the pump photons is uniformly perpendicularly incident on the metasurfaces with a power of 2 mW and a 2 $\mu$m (diameter) diffraction-limited spot\cite{Marino2019}. The normalized SFG efficiency is independent on the incident intensities and relies only on the resonant nanostructure. The estimated photon-pair rates of SPDC process are shown in Fig. 4. Similarly as the SFG efficiency, the prominent SPDC enhancement are located near the wavelengths of the quasi-BIC resonances for both the generated photons. Such strong enhancement of photon-pair rate in the designed metasurfaces results from the resonant enhancement effect of the dual quasi-BICs, because the efficient photon-pair generation requires a high density of states at both signal and idler wavelengths. In particular, an important feature of the spectral brightness in Fig. 4(a) is the giant enhancement of SPDC rate within a narrow resonance bandwidth. It is the direct consequence of the ultrahigh-$Q$ quasi-BIC resonances with narrow transmission spectra of signal or idler beams. To further confirm the resonant enhancement of photon generation, the photon-pair rate as a function of signal wavelength $\lambda_{\text{s}}$ is shown in Fig. 4(b), with a fixed pump wavelength of $\lambda_{\text{p}}=760$ nm. The photon-pair generation rate curve shows the peak with the high value of $1.05\times10^{4}$ Hz. It highlights the significant contribution of quasi-BIC resonances in enhancing density of states for the vacuum field fluctuations in SPDC process. Compared with that of Mie resonant nanostructures composed of isolated AlGaAs nanoantennas with similar pump intensities\cite{Marino2019}, the SPDC rates in the proposed metasurface driven by dual quasi-BICs is increased by about two orders of magnitude. 

\begin{figure}[htbp]
	\centering
	\includegraphics% Here is how to import EPS art
	[scale=0.35]{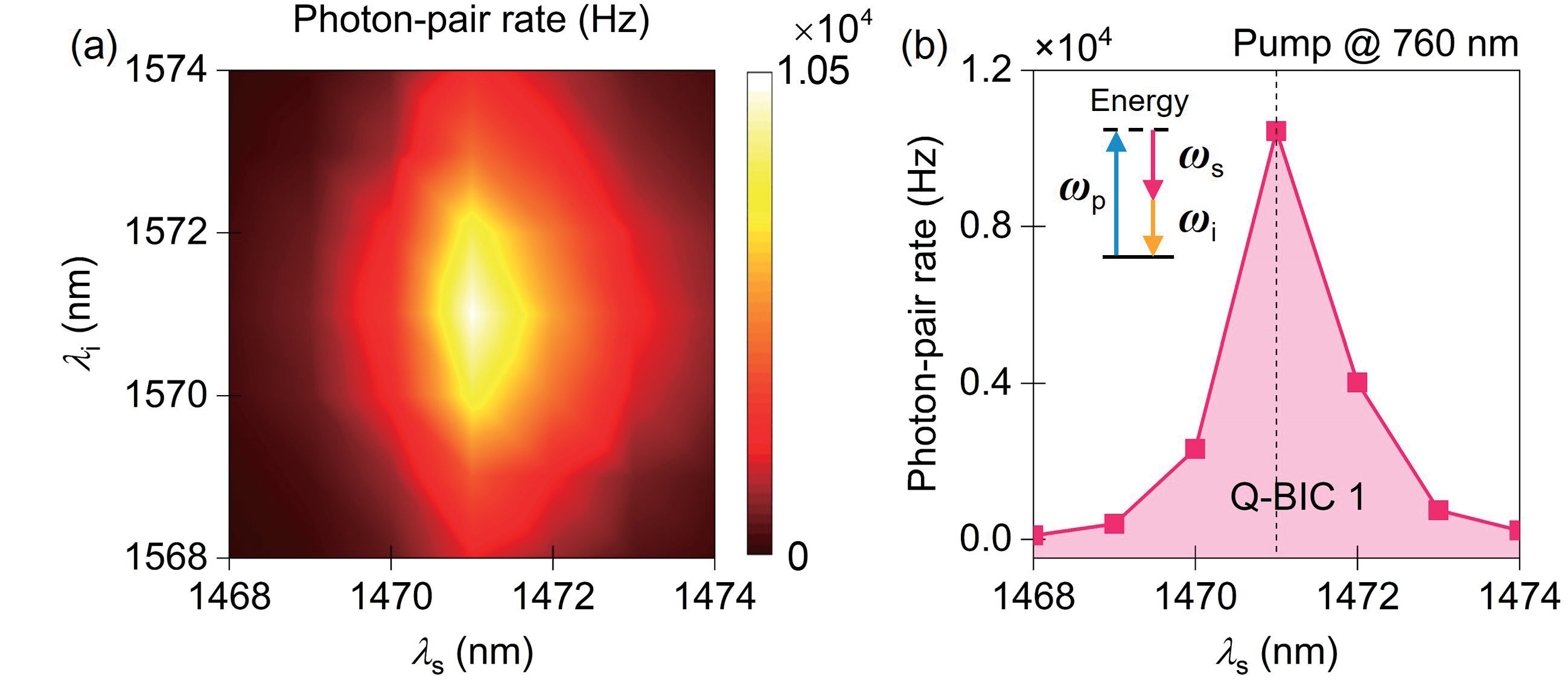}
	\caption{\label{fig4} Enhanced photon-pair generation rate of SPDC in the dual quasi-BIC metasurface. The SPDC photon-pair rate as a function of (a) the wavelengths of signal and idler beams and (b) the signal wavelength when the pump wavelength is constant at $\lambda_{\text{p}}=760$ nm. }
\end{figure}

We then explore the dependence of the SFG efficiency and photon pair rates on the rotation angle of AlGaAs ellipse cylinders in the proposed resonant metasurface. As shown in Fig. 5, both the nonlinear conversion efficiencies in SFG and SPDC processes show dramatically decline as rotation angle $\theta$ gradually increases. The rapid decline tendency can be explained by the reduced $Q$ factors of the quasi-BICs with larger $\theta$. Because the rotation angle of AlGaAs nano-resonators controls the mode coupling and radiation loss of the metasurface to the far field, it is able to engineer the $Q$ factor of the two quasi-BICs. As depicted in Fig. S1(b), the $Q$ factor dramatically reduces with increasing $\theta$. This also agrees well with the broaden transmission dips in Fig. 2(a). Accordingly, the capability of electric field confinement of the quasi-BIC resonances which are directly related to $Q$ factors are weakened for reduced nonlinear efficiency. Note that both the resonant enhancement of SFG and SPDC processes are the combined results of the two quasi-BICs in the proposed metasurface, the SFG efficiency and the photon-pair rate of SPDC are jointly determined by the $Q$ factors of the two resonances. Thus, the sensitivity of $Q$ factor to the rotation angle $\theta$ brings the flexibility to engineer the quantum photon generation rate. 

\begin{figure}[htbp]
	\centering
	\includegraphics% Here is how to import EPS art
	[scale=0.4]{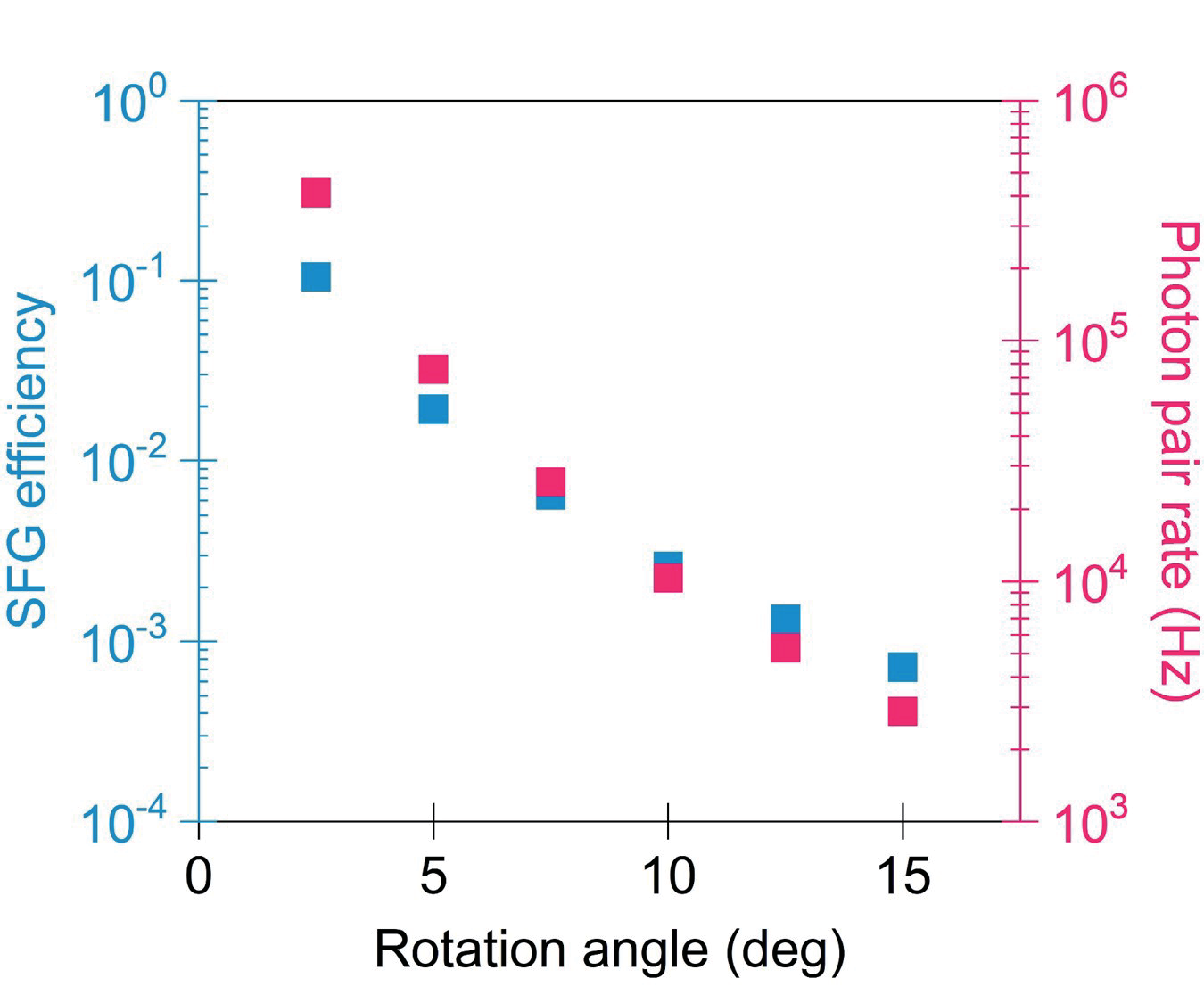}
	\caption{\label{fig5} SFG efficiency and photon-pair rate of SPDC at quasi-BIC resonances as a function of the rotation angle of the metasurface.}
\end{figure}

Finally, we provide the photo-pair generation rate at different observation angles of the signal and idler waves in Fig. 6. The generated photon pairs at the dual quasi-BIC resonances of the metasurfaces are considered, with the signal and idler at wavelengths of 1471 nm and 1571 nm and the pump at 760nm, respectively. The observation angles of signal and idler denoted as $\theta_{\text{s}}$ and $\theta_{\text{i}}$ respectively, are defined with respect to the normal direction of the metasurface. The emitted signal and idler photon rates are centered around the normal direction, as shown in Fig. 6(a). Considering the resonant wavelengths of the quasi-BICs in the metasurface are sensitive to the incident angle of the fundamental waves due to the high-$Q$ features, in turn, the photo-pair generation of SPDC process exhibit narrowband brightness within a few degrees of the observation angles. The photon pair rate keeps at a high value around $10^{4}$ Hz within observation angle $\theta_{\text{i}}$ from $-2^{\circ}$ to $2^{\circ}$ and $\theta_{\text{s}}$ from $-0.5^{\circ}$ to $0.5^{\circ}$, respectively, and dramatically reduce beyond these observation scope. This tendency can also be illustrated in Fig. 6(b) by the dependence of the generated photon rates on the observation angles $\theta_{\text{s}}$. In the proposed metasurface, the optimized observation angle for the maximum radiation power is normal to the surface, and the wider observation angle $\theta_{\text{i}}$ relative to $\theta_{\text{s}}$ can be explained by the smaller $Q$ factor and broader transmission linewidth of Q-BIC 2 for idler wave than those of Q-BIC 1 for signal wave. 

\begin{figure}[htbp]
	\centering
	\includegraphics% Here is how to import EPS art
	[scale=0.35]{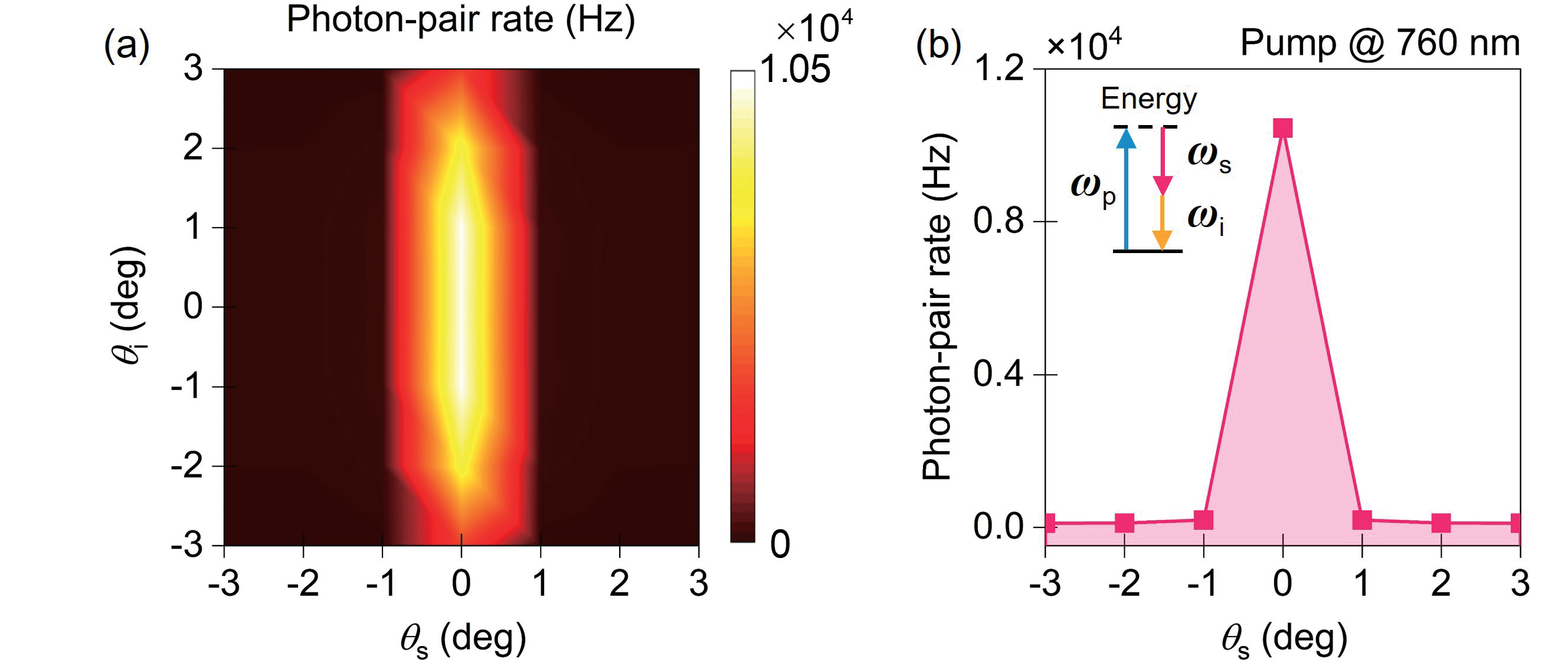}
	\caption{\label{fig6} Photon-pair generation rate of SPDC at the quasi-BIC resonances in the metasurface at different observation angles, with the $\lambda_{\text{s}}=1471$ nm, $\lambda_{\text{i}}=1571$ nm, and $\lambda_{\text{p}}=760$ nm. The SPDC photon-pair rates as a function of (a) the signal and idler wave observation angles ($\theta_{\text{s}}$ and $\theta_{\text{i}}$) and (b) the observation angles of signal beam when the idler beam is observed with $\theta_{\text{i}}=0^{\circ}$ at $\lambda_{\text{p}}=760$ nm. }
\end{figure}

\section{\label{sec4}Conclusions}

In conclusion, we have theoretically investigated the efficient photon-pair generation via SPDC from a nonlinear dielectric metasurface. In the simple metasurface design using AlGaAs elliptical nano-cylinders as building blocks, the two quasi-BIC resonances with ultrahigh-$Q$ factors are employed to generate efficient nonlinear processes. By coinciding the two resonances with the generated signal and idler frequencies, the electric field energy in the nonlinear material is significantly enhanced. Due to the strongly enhanced nonlinear light-matter interactions, the reverse classical nonlinear process of SFG is substantially boosted with the efficiency up to $2.64\times10^{-3}$. According to the quantum-classical correspondence relation, the generation rate of photon pairs is remarkably increased to $1.05\times10^{4}$ Hz, which is two orders of magnitude larger than that in the Mie resonant AlGaAs nanoantennas. Enhanced emission of photon pairs is produced only in a narrow wavelength range due to the nonlocal feature of quasi-BIC resonances. Furthermore, the generation rates of photon pair via SPDC could be controlled through engineering $Q$ factors via adjusting rotation angles of nano-cylinders. We also demonstrate that the generated signal and idler photons are emitted mainly concentrated in the normal direction with respect to the metasurface. These results will help in the design of metasurface for the generation of high-quality entangled photon pairs, fueling the development of the nanoscale sources of quantum light. Finally, it is worth while to note that the strategy and metasurface design here can be extended for highly selective enhancement of SFG and SPDC processes at alternative wavelengths via the scalability of Maxwell's equations, thus enabling multifrequency quantum states including cluster state generation from a single metasurface chip. $\chi^{(2)}$ two-dimensional materials can be integrated for these nonlinear and quantum phenomena as well\cite{Trovatello2020, DinparastiSaleh2018, Yuan2019, Yao2020, Kim2020, Bernhardt2020, Loechner2021, Liuzj2021}.

\begin{acknowledgments}	
This work was supported by the National Natural Science Foundation of China (Grants No. 12304420, 12264028, 12364045, 12364049, and 12104105), the Natural Science Foundation of Jiangxi Province (Grants No. 20232BAB201040 and 20232BAB211025), the Young Elite Scientists Sponsorship Program by JXAST (Grant No. 2023QT11), and the Innovation Fund for Graduate Students of Jiangxi Province (Grant No. YC2023-S028).	

T. L. and M. Q. contributed equally to this work. M. Q. has received a Ph.D. degree in physics from Nanchang University and is currently starting her teaching \& research career at Nanchang Institute of Science and Technology.

\end{acknowledgments}

%\bibliography{Ref}% Produces the bibliography via BibTeX.

\end{document}